# Hybrid Electro-Optically Modulated Microcombs

Pascal Del'Haye[1*], Scott B. Papp[1], Scott A. Diddams[1†]

[1]*National Institute of Standards and Technology (NIST), Boulder, CO 80302, USA*

**Optical frequency combs based on mode-locked lasers have proven to be invaluable tools for a wide range of applications in precision spectroscopy and metrology. A novel principle of optical frequency comb generation in whispering-gallery mode microresonators ("microcombs") has been developed recently, which represents a promising route towards chip-level integration and out-of-the-lab use of these devices. Presently, two families of microcombs have been demonstrated: combs with electronically detectable mode spacing that can be directly stabilized, and broadband combs with up to octave-spanning spectra but mode spacings beyond electronic detection limits. However, it has not yet been possible to achieve these two key requirements simultaneously, as will be critical for most microcomb applications. Here we present a key step to overcome this problem by interleaving an electro-optic comb with the spectrum from a parametric microcomb. This allows, for the first time, direct control and stabilization of a microcomb spectrum with large mode spacing (>140 GHz) with no need for an additional mode-locked laser frequency comb. The attained residual 1-second-instability of the microcomb comb spacing is $10^{-15}$, with a microwave reference limited absolute instability of $10^{-12}$ at a 140 GHz mode spacing.**

Microresonator-based optical frequency combs ("microcombs") are a promising candidate to combine precision metrology applications with miniaturized and chip-scale microphotonic devices [1, 2]. High conversion efficiency and large mode spacings make optical microcombs ideal for applications that require high power per comb line and the possibility to optically resolve single comb modes. These attributes are especially interesting for direct comb spectroscopy, astrophysical spectrometer calibration [3], arbitrary optical waveform generation [4] and general telecommunication applications. After the first demonstration in 2007, the rapidly growing field of microcombs has led to the development of a variety of systems in different materials, including fused silica toroids [1], disks and microrods [5, 6], crystalline resonators made of calcium fluoride [7, 8] and magnesium fluoride [9], as well as fully integrated silicon photonic devices [10, 11] and high-index silica resonators [12, 13]. A common feature of all these microcomb systems is that the comb itself is directly generated via nonlinear frequency conversion (four-wave mixing) from a single continuous-wave pump laser. In such energy-conserving processes, two symmetric sidebands are generated by converting two pump photons (frequency $\omega_p$) into a signal ($\omega_s$) and idler photon ($\omega_i$) obeying $2\omega_p = \omega_s + \omega_i$. The frequency spacing between pump photon and the sidebands is associated with an integer multiple of the resonator's free spectral range. Recent studies have suggested that one possibility to ensure low-noise microcomb operation is the initial generation of sidebands with a separation close to a single free spectral range with respect to the pump frequency. This regime, however, requires strong accumulated anomalous dispersion between adjacent modes, which is intrinsically given in smaller resonators as a result of larger mode spacings [14]. However, the mode spacings in these smaller resonators are not amenable to direct electronic measurements, which is a key prerequisite for frequency stabilization schemes. Here we present a hybrid approach that combines low-noise microcomb generation in small resonators with electro-optic sideband generation for bridging the gaps between microcomb modes. This scheme enables the first direct referencing of a microcomb mode spacing >100 GHz to a microwave frequency standard and is a key prerequisite for metrological applications of microcombs. A similar approach has been shown for the stabilization of dual mode lasers in previous work, which allowed for mode separations around 100 GHz with a relative frequency instability of $10^{-11}$ [15]. The attained mode spacing stability is a factor of five better than the best reported stability in smaller microresonators with directly accessible mode spacing [6], and is commensurate with state-of-the-art frequency standards that could employ such a comb in the development of optical clocks [16, 17]. In addition, we present a scheme in which we stabilize the resonator mode spacing to an exact harmonic of the electro-optic sideband frequency. By using different sub-harmonic frequencies of the electro-optical modulator, this enables the generation of optical frequency combs with nearly arbitrary mode spacings from a single comb generator. This is particularly interesting for astrophysical spectrometer calibration and telecommunication applications, in which the comb spacing can be adjusted to the resolution of the employed optical devices. Moreover, this scheme would provide stable optical signals for high-purity microwave- and terahertz-frequency synthesis [15] as a result of the interaction of a high number of microcomb modes. In particular mode spacings in the terahertz range could be achieved by implementing advanced modulation techniques using $LiTaO_3$ crystals [18] or cavity enhanced $LiNbO_3$ electro optic modulation [19].

**Results**

The experimental setup and scheme of a hybrid electro-optically modulated microcomb generator are shown in Figure 1. A tunable external cavity diode laser at a wavelength around 1556 nm is used to send light into a 230-μm-radius fused silica microrod resonator ([6], Fig. 1b,c) with an optical quality factor of Q = 4.1×10$^8$. Pumping with 250 mW optical power, far above the parametric comb generation threshold (~500 μW), generates a broad optical comb with ≈ 143 GHz mode spacing. Part of the microcomb is sent to a Mach-Zehnder-type lithium niobate electro-optic intensity modulator (EOM), which is driven at $f_{eom}$ ≈ 23 GHz and generates several orders of optical sidebands around each comb mode (cf. Fig. 1d). The EOM sidebands from two adjacent microcomb modes meet in between the microcomb modes and generate a low-frequency beat note $f_{beat}$ [15]. The actual microcomb mode spacing $f_{rep}$ directly relates to this beat note via $f_{rep}$=n × $f_{eom}$ ± $f_{beat}$, where n is the integer number of EOM sidebands between two microcomb modes. Subsequently, the low-frequency beat note is used to stabilize the microcomb mode spacing to a frequency standard via a phase-locked loop. Here, the launched optical power into the resonator serves as actuator to control the mode spacing via a fast thermal effect [20]. The power is controlled via the pump-diode current of an erbium-doped fiber amplifier. An additional comb stabilization scheme, in which the microcomb spacing is phase locked to an integer harmonic of the EOM frequency, is presented later in Figure 4.

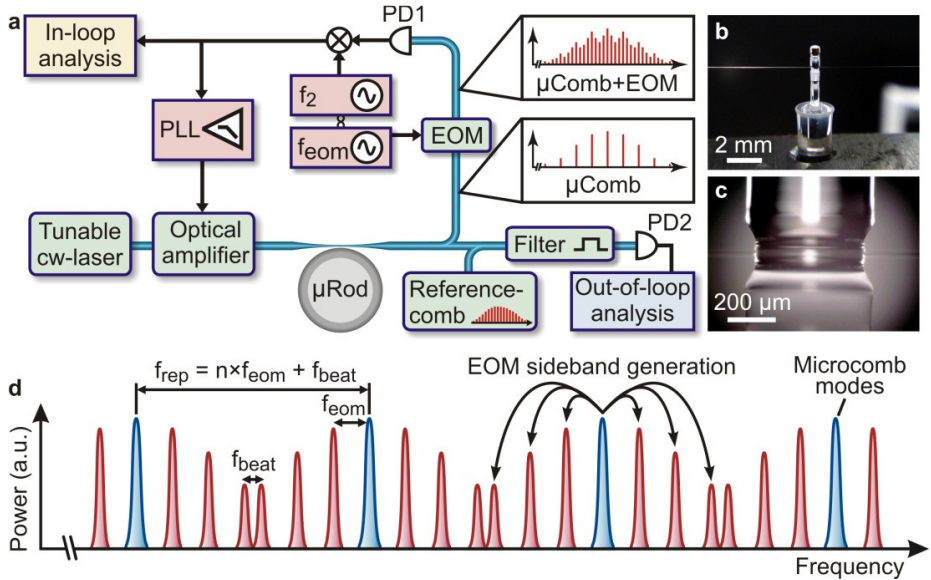

**Figure 1: Experimental scheme of electro-optically modulated microcombs**. (**a**) Setup for generation and stabilization of microcombs with electro-optical modulation. PD=photodiode, EOM=electro-optic modulator, PLL=phase-locked loop. (**b,c**) Photograph and microscope image of the employed whispering-gallery mode Microrod resonator with a radius of 230 micrometers (143 GHz mode spacing). (**d**) Scheme of electro-optically generated sidebands around each microcomb mode. The low-frequency beat note $f_{beat}$ between the sidebands is used to control and stabilize the comb spacing.

Measurements of unknown frequencies with conventional optical combs often require an independent additional measurement in order to determine the exact mode index of the comb line that generates a particular beat note. The hybrid EOM microcomb, in contrast, offers the unique opportunity to slightly vary the EOM frequency and monitor the beat-note shift in order to determine the exact number of EOM sidebands by which an optical signal is separated from a microcomb mode. Thus, an exact frequency measurement only requires the knowledge of the frequency of the closest microcomb mode, which can be easily measured with an optical spectrum analyzer because of the wide mode spacing.

The optical spectrum of a 143-GHz-spaced microcomb with and without electro-optic modulation is shown in Fig. 2. Three orders of electro-optic sidebands with a spacing of 23 GHz are generated around each microcomb mode. The beat note between the third-order sidebands from two adjacent microcomb modes is at a sufficiently low frequency to measure and stabilize the microcomb spacing. In order to change the EOM sideband spacing we can adjust the microwave synthesizer frequency to values between 21 GHz and 27 GHz (this range is limited by the amplifier that generates the +27 dBm drive signal for the EOM).

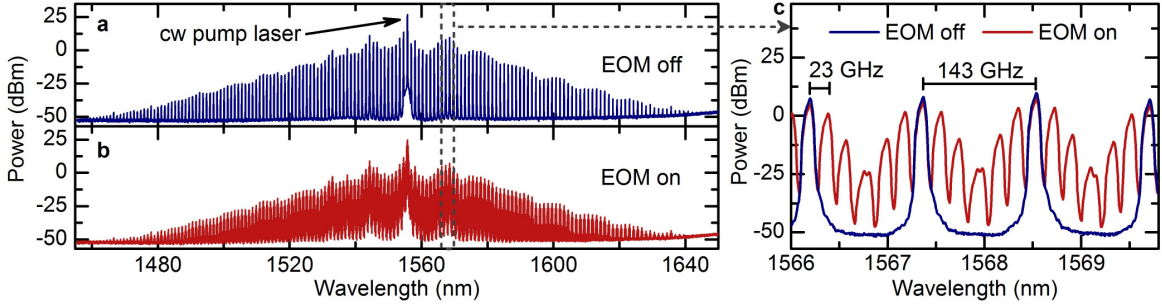

**Figure 2: Optical microcomb spectra with and without electro-optic modulation sidebands**. (**a**) Optical comb from a microrod resonator with 143 GHz mode spacing. The launched power into the tapered optical fiber is ~250 mW. (**b**) Same comb spectrum filled with sidebands from the electro-optic modulator. (**c**) Zoom into the combined spectra with and without electro-optic modulation. Both the 143 GHz microcomb spacing as well as the 23 GHz electro-optic modulation sidebands are visible. The third-order electro-optic sidebands are nearly overlapping and cannot be distinguished by the optical spectrum analyzer (resolution ~3 GHz). The beat note between these central sidebands is used for measurement and stabilization of the microcomb mode spacing.

The microcomb spectrum of Fig. 2 is used to test the viability of the EOM-assisted mode-spacing stabilization scheme. The corresponding mode-spacing Allan deviations for the free-running as well as the stabilized microcomb are depicted in Fig. 3. Activating the stabilization loop via the pump power reduces the 1-second-instability of the comb spacing from $2.5\times10^{-8}$ (free running) down to $1.2\times10^{-15}$ (in-loop). A slightly increased Allan deviation of $2.9\times10^{-15}$ is observed for a mode-spacing stabilization to an harmonic of the EOM frequency, which is attributed to a different electronic signal-processing, as described in Fig. 4. In order to verify that the stabilization of the beat note between the EOM sidebands indeed stabilizes the actual microcomb spectrum, we have performed an independent out-of-loop measurement against a reference frequency comb (cf. Fig. 1a). The employed reference comb is a mode-locked and fully referenced Erbium-doped fiber laser with a repetition rate of 250 MHz. In the out-of-loop measurement we generate two beat notes of adjacent microcomb modes with neighboring fiber-laser comb modes. The influence of carrier-envelope frequency fluctuations is eliminated by generating the difference frequency of the two beat notes, which solely depends on the fluctuations of the two combs' mode spacings. This difference frequency is subsequently sent to a phase-noise analyzer in order to determine the fractional frequency stability. The out-of-loop measurements show a 1-second-instability of $1\times10^{-12}$, which is only limited by the stability of the employed microwave synthesizers that is also imprinted on the reference comb; see Fig. 3a. The electronic spectrum of the generated difference frequency between two beat notes against the fiber reference comb is shown in Fig. 3c, along with a spectrum of the in-loop beat note signal in Fig. 3b. In an additional measurement we have separately measured the residual instability introduced by the electro-optic modulation, photo-detection, down-mixing and amplification to be $3.5\times10^{-16}$ at 1 second gate time. This implies that the out-of-loop-stability could be vastly improved with better microwave references.

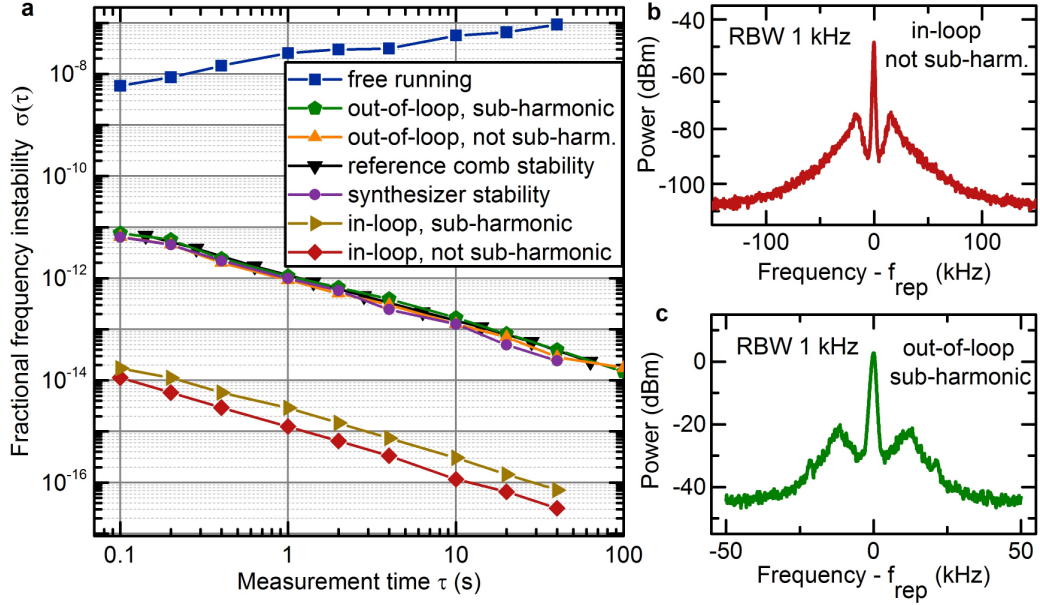

**Figure 3: Frequency stability of the free-running and stabilized microcomb.** (**a**) Allan-deviation measurements with different gate times. The fractional (absolute) one-second-instabilities of the microcomb mode spacing are $\sigma = 2.5\times10^{-8}$ (3.6 kHz) in the free-running case, $\sigma = 1\times10^{-12}$ (140 mHz) for the out-of-loop measurements, and $\sigma = 1.2\times10^{-15}$ (170 µHz) for the in-loop measurement. Note that the out-of-loop stabilities are limited by the synthesizer that produces $f_{eom}$ (shown by the purple circles), which also limits the stability of the reference comb (black upside-down triangles). All fractional frequency instabilities are given at the microresonator mode spacing frequency of 143 GHz. (**b**) Measurement of the stabilized electronic spectrum of the in-loop beat note $f_{beat}$ in Fig. 1d. (**c**) Out-of-loop mode-spacing beat note against a fiber-laser reference comb. This beat note is generated by beating two adjacent microcomb lines against a stabilized reference comb, followed by generation of the difference frequency of these two beats. This beat note depends only on the microcomb mode spacing and not the offset frequency.

For certain applications, e.g. for an evenly distributed channel spectrum in telecommunications or arbitrary waveform generation, it is advantageous to generate a continuous comb, which means that the microcomb mode spacing be an exact harmonic of the electro-optic modulation frequency. However, this requires stabilizing the beat between the central EOM sidebands ($f_{beat}$ in Fig. 1d) to zero frequency, which is challenging due to low-frequency noise in the laser system and photodetector. We circumvent this difficulty by generating a higher-frequency beat note ($f_{beat1}$ in Fig. 4a) along with three times the electro-optic modulation frequency $f_{eom}$. The microcomb spacing is locked to a harmonic of $f_{eom}$ when $f_{beat1}-3\times f_{eom}=0$. In this scheme the difference frequency $f_{beat1}-3\times f_{eom}$ is used as an error signal for a phase-locked loop, which links the microcomb spacing to exactly 6 times the EOM sideband frequency. Figure 4b shows the measured signal $\Delta f_{rep}=f_{beat1}-3\times f_{eom}$, which directly corresponds to the microcomb mode spacing change $\Delta f_{rep}$, as a function of launched optical power. The total tuning range of the microcomb spacing is around 1.5 MHz, with a slope of approximately -6 kHz/mW. Attained instabilities of this sub-harmonic locking scheme are plotted in Fig 3a (traces labeled with "sub-harmonic" in the legend), and show only slightly worse values compared to the non-sub-harmonic case.

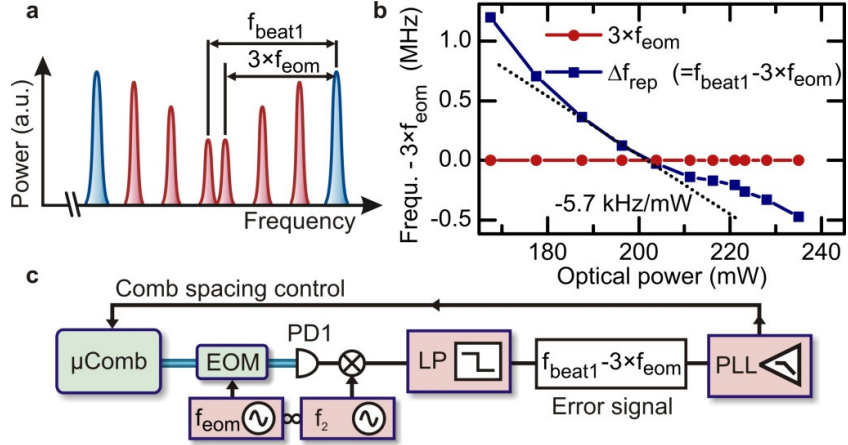

**Figure 4: Sub-harmonic stabilization of a microcomb**. (**a**) Principle of generation and stabilization of eletro-optic sidebands at a sub-harmonic frequency of the microcomb spacing. A fast photodiode (PD1 in panel c) simultaneously detects $f_{beat1}$ and $3\times f_{eom}$. The subsequently generated difference frequency $f_{beat1}-3\times f_{eom}$ is used to stabilize the microcomb spectrum at exactly six times the electro-optic modulation frequency. This scheme provides a continuous optical comb with a spacing of $f_{eom}$. The corresponding mode spacing stability is shown in Fig 3a. (**b**) Crossing of $f_{beat1}$ and $3\times f_{eom}$ when changing the launched optical power. The sub-harmonic lock-point corresponds to $f_{beat1}-3\times f_{eom}=0$. The total mode-spacing tuning range is around 1.5 MHz, with a slope of -5.7 kHz/mW. (**c**) Setup for the sub-harmonic comb stabilization. EOM = electro-optic modulator, PD = photodiode, LP = low-pass, PLL = phase-locked loop.

## Discussion

We introduced a novel scheme for measurement and stabilization of the mode spacing in widely spaced optical frequency combs via electro-optic modulation. This stabilization scheme is expected to be a key prerequisite for broadband and low-noise microcomb generation for metrology applications, as well as for integrated micro- and nano-photonic devices. In addition, this scheme could provide stable optical signals for high-purity terahertz-frequency synthesis [15]. The presented results provide a route to overcome bandwidth limitations in larger microcomb generators while maintaining the ability to directly access and stabilize the comb spacing. In addition, this scheme provides a means to adapt optical frequency comb spectra to different applications by controlled subdivision of the initial comb spacing. While here we demonstrate stabilization of a 140-GHz-microcomb, we envision that with improved modulation techniques [18] our approach should work up to mode spacings in the THz range. In addition, future work could potentially include integration of the electro-optic modulator directly with a chip-based microcomb generator by use of technology demonstrated in related silicon photonic devices [21].

**Acknowledgements:** The authors thank Gabe Ycas for providing the fiber laser reference comb. This work is supported by NIST and the DARPA QuASAR program. PD thanks the Humboldt Foundation for support. This paper is a contribution of NIST and is not subject to copyright in the United States.

[*]pascal.delhaye@gmx.de

[†]scott.diddams@nist.gov